\documentclass[12pt]{article}%
\usepackage{amsfonts}
\usepackage{amsmath}
\usepackage{geometry}
\usepackage{amssymb}
\usepackage{graphicx}
\usepackage[onehalfspacing]{setspace}
\usepackage{hyperref}%
\providecommand{\U}[1]{\protect\rule{.1in}{.1in}}
\begin{document}

\title{\textbf{Mass Formula of a Five-dimensional Almost-BPS Supergravity Soliton
with a Magnetic \textquotedblleft Bolt\textquotedblright\medskip}}
\author{\textbf{Patrick A. Haas\medskip}\\Department of Physics and Astronomy\\University of Southern California\\Los Angeles, CA 90089, USA\\\medskip\\phaas@usc.edu}
\date{}
\maketitle

\begin{abstract}
\noindent\thispagestyle{empty}We derive the Smarr formula for a
five-dimensional spacetime which has a magnetic \textquotedblleft
bolt\textquotedblright\ in its center and is asymptotically $%
%TCIMACRO{\U{211d} }%
%BeginExpansion
\mathbb{R}
%EndExpansion
^{1,3}\times S^{1}$. Supersymmetry -- and so the BPS-bound -- is broken by the
holonomy. We show how each topological feature of a space-like hypersurface
enters the mass formula and which ones in particular give rise to the
violation of the BPS-bound.\newpage\setcounter{page}{1}

\end{abstract}
\tableofcontents

\section{Introduction}

Finding a proper quantum description of black holes is of major interest to
solve profound problems surrounding the classical picture, like singularities
and the \textquotedblleft information paradox\textquotedblright.

One possible way is the in 2002 by Samir Mathur proposed \textquotedblleft
fuzzball\textquotedblright\ program \cite{Mathur - Fuzzball Review} within the
framework of string theory whose supergravity limit is smooth, horizonless and
asymptotically flat solutions representing time-independent solitons.

In recent works it has been shown that horizonless solitonic solutions of
supergravity can indeed be constructed purely by means of nontrivial topology;
the Smarr formula has been derived in each case by means of the Komar integral
formalism over cohomology
\cite{Gibbons:2013tqa,Haas:2014,Vercnocke:2015,Kunduri}. One important result
is the role of Chern-Simons terms to only support the topological nature of
the integral.

In this work we recall the five-dimensional results of \cite{Gibbons:2013tqa}
but impose a different choice of boundary conditions; in particular, we assume
a spacetime which has a magnetic \textquotedblleft bolt\textquotedblright\ in
its center and is asymptotically $%
%TCIMACRO{\U{211d} }%
%BeginExpansion
\mathbb{R}
%EndExpansion
^{1,3}\times S^{1}$ -- the fourth spatial dimension is periodic, and
three-dimensional space is rotating along it with nonzero angular momentum
even at infinity. To construct it, we define in the fashion of
\cite{Warner:2009} a four-dimensional Ricci-flat base space which carries
Euclidean Schwarzschild metric and magnetic flux from a \textquotedblleft
floating brane\textquotedblright\ ansatz \cite{FloatingBranes2010} for the
Maxwell fields.

The supersymmetry conditions require that the curvature tensor be either
self-dual or anti-self-dual and tell how this duality has to be correlated
with the one of the Maxwell-fields. Since the rotation group in
four-dimensional space decomposes like $SO\left(  4\right)  =SU\left(
2\right)  _{\text{self-dual}}\times SU\left(  2\right)
_{\text{anti-self-dual}}$, only one half of the Killing-spinors would
\textquotedblleft feel\textquotedblright\ space's holonomy and the other half
flat space. In simple examples, this half-flatness, and the preservation or
breaking of supersymmetry can be easily arranged by just changing a sign in
the duality of the fields \cite{GoldsteinKatmadas, BenaWarnerNonBPS,
BenaWarnerMultiCenter}.

However, the curvature tensor for the Schwarzschild bolt is neither self-dual
nor anti-self-dual, so our system is manifestly non-BPS; but, because the
Schwarzschild geometry is still Ricci-flat, the (almost-)BPS-equations are
still satisfied \cite{Warner:2009,FloatingBranes2010}, and hence provides a
ground for more general solutions. It is also for the above stated
circumstances that one speaks in the present context of \textquotedblleft
almost-BPS\textquotedblright-solutions.

The idea here is to derive a Smarr formula based on the Komar-integral
formalism in the sense of \cite{Gibbons:2013tqa} for the present boundary
conditions and for non-BPS solutions and see in how far the arising mass
components -- especially the BPS-bound breaking terms -- relate explicitely to
the structure of the given spacetime, particularly space's boundary.

Finally, expressions for the five-dimensional masses will be compared to those
from \cite{Warner:2009}.

\section{Preliminaries}

\subsection{The five-dimensional Supergravity action and equations of motion}

The bosonic action in five dimensions \cite{Bena&Warner:2007} is:%
\begin{equation}
S=\int\left(  \star_{5}R-Q_{IJ}dX^{I}\wedge\star_{5}dX^{J}-Q_{IJ}F^{I}%
\wedge\star_{5}F^{J}-\tfrac{1}{6}C_{IJK}F^{I}\wedge F^{J}\wedge A^{K}\right)
, \label{Action_5D}%
\end{equation}
where $C_{IJK}=\left\vert \epsilon_{IJK}\right\vert $, $X^{I}$ ($I=1,2,3$) are
scalar fields and the $A^{I}$ are Maxwell fields.

We may consider this to arise from the on $T^{6}$ reduced eleven-dimensional
theory, where the scalars come from the metric coefficients of the
compactified dimensions,%
\begin{equation}
ds_{11}^{2}=ds_{5}^{2}+\left(  \tfrac{Z_{2}Z_{3}}{Z_{1}^{2}}\right)
^{\frac{1}{3}}\left(  dx_{5}^{2}+dx_{6}^{2}\right)  +\left(  \tfrac{Z_{1}%
Z_{3}}{Z_{2}^{2}}\right)  ^{\frac{1}{3}}\left(  dx_{7}^{2}+dx_{8}^{2}\right)
+\left(  \tfrac{Z_{1}Z_{2}}{Z_{3}^{2}}\right)  ^{\frac{1}{3}}\left(
dx_{9}^{2}+dx_{10}^{2}\right)  , \label{Metric_11D}%
\end{equation}
with the reparametrization,%
\begin{equation}
X^{1}=\left(  \tfrac{Z_{2}Z_{3}}{Z_{1}^{2}}\right)  ^{\frac{1}{3}},\text{
}X^{2}=\left(  \tfrac{Z_{1}Z_{3}}{Z_{2}^{2}}\right)  ^{\frac{1}{3}},\text{
}X^{3}=\left(  \tfrac{Z_{1}Z_{2}}{Z_{3}^{2}}\right)  ^{\frac{1}{3}},
\label{Reparameterization}%
\end{equation}
to fulfill the constraint $X^{1}X^{2}X^{3}=1.$

Moreover, there is a metric for the kinetic terms,%
\begin{equation}
Q_{IJ}=\tfrac{1}{2}\text{diag}\left(  \left(  \tfrac{1}{X^{1}}\right)
^{2},\left(  \tfrac{1}{X^{2}}\right)  ^{2},\left(  \tfrac{1}{X^{3}}\right)
^{2}\right)  , \label{Metric_kinetic terms}%
\end{equation}
and the duals of the field strengths, $F^{I}=dA^{I}$, are then given by%
\begin{equation}
G_{I}=Q_{IJ}\left(  \star_{5}F^{J}\right)  . \label{G_5D}%
\end{equation}

The Einstein equations \cite{Gibbons:2013tqa} are%
\begin{equation}
R_{\mu\nu}=Q_{IJ}\left(  F_{\mu\rho}^{I}F_{\nu}^{J\rho}-\tfrac{1}{6}g_{\mu\nu
}F_{\rho\sigma}^{I}F^{J\rho\sigma}+\partial_{\mu}X^{I}\partial_{\nu}%
X^{J}\right)  , \label{Einstein}%
\end{equation}
and the Maxwell equations,%
\begin{equation}
\nabla_{\rho}\left(  Q_{IJ}F_{\text{ \ \ \ }\mu}^{J\rho}\right)  =J_{I\mu
}^{CS}, \label{Maxwell}%
\end{equation}
with the five-dimensional Chern-Simons 1-form current\footnote{The Levi-Civita
tensor for curved spacetime is related to the Levi-Civita symbol of Minkowski
spacetime like $\bar{\epsilon}^{\mu_{1}...\mu_{5}}=\left(  -g\right)
^{-\frac{1}{2}}\epsilon^{\mu_{1}...\mu_{5}}\Leftrightarrow\bar{\epsilon}%
_{\mu_{1}...\mu_{5}}=\left(  -g\right)  ^{\frac{1}{2}}\epsilon_{\mu_{1}%
...\mu_{5}}$ with the convention $\epsilon_{01234}=1$.},%
\begin{equation}
J_{I\mu}^{CS}=\tfrac{1}{16}C_{IJK}\bar{\epsilon}_{\mu\rho\sigma\kappa\lambda
}F^{J\rho\sigma}F^{K\kappa\lambda}. \label{CS-current}%
\end{equation}
Moreover, this can be expressed in terms of the dual field strengths,%
\begin{equation}
dG_{I}=\tfrac{1}{4}C_{IJK}F^{J}\wedge F^{K}. \label{dG}%
\end{equation}

Eq. $\left(  \ref{Einstein}\right)  $ can be rewritten such that the RHS is
free of any trace terms,%
\begin{equation}
R_{\mu\nu}=Q_{IJ}\left(  \tfrac{2}{3}F_{\mu\rho}^{I}F_{\nu}^{J\rho}%
+\partial_{\mu}X^{I}\partial_{\nu}X^{J}\right)  +\tfrac{1}{6}Q^{IJ}G_{I\mu
\rho\sigma}G_{J\nu}^{\text{ \ \ }\rho\sigma}, \label{Einstein_rearranged_1}%
\end{equation}
especially since this form is much more helpful for the derivation of the
Komar mass formula.

\subsection{Invariances and the Komar integral}

We assume the metric to have a time-like Killing vector, $K^{\mu}$, and can
hence write the five-dimensional mass formula in terms of a Komar integral in
five dimensions,%
\begin{equation}
M=\tfrac{3}{32\pi G_{5}}\int_{X^{3}}\star_{5}dK, \label{Komar integral}%
\end{equation}
where $X^{3}$ is the 3-boundry of the five-dimensional spacetime. Smoothness
of spatial sections, $\Sigma_{4}$, allows in virtue of properties of the
Killing vector to rewrite this formula as an integral over such by $X^{3}$
bound space-like hypersurfaces:%
\begin{equation}
M=\tfrac{3}{32\pi G_{5}}\int_{X^{3}}\star_{5}dK=\tfrac{3}{16\pi G_{5}}%
\int_{\Sigma_{4}}K^{\mu}R_{\mu\nu}d\Sigma^{\nu}.
\end{equation}

Assuming, furthermore, that the matter fields have the symmetries of the
metric, means them to be invariant under the Lie-derivative along the Killing
vector, $K$,%
\begin{equation}
\mathcal{L}_{K}F=0=\mathcal{L}_{K}G,\label{Invariance}%
\end{equation}
where $\mathcal{L}_{K}$ is the corresponding Lie derivative, we get, in the
same fashion as in \cite{Gibbons:2013tqa}, the equations,%
\begin{equation}
0=d\left(  i_{K}F\right)  \Leftrightarrow i_{K}F^{I}=d\lambda^{I}\text{ and
}i_{K}G_{I}=d\Lambda_{I}-\tfrac{1}{2}C_{IJK}\lambda^{J}F^{K}+H_{I}^{\left(
2\right)  },\label{F_G_Killing}%
\end{equation}
where $\lambda^{I}$ are magnetostatic potentials of the $G_{I}$ and
electrostatic potentials of the $F^{I}$, respectively; $\Lambda_{I}$ are
globally defined 1-forms and $H_{I}^{\left(  2\right)  }\in H^{2}\left(
\mathcal{M}_{5}\right)  $ closed but not exact 2-forms.

With $\left(  \ref{F_G_Killing}\right)  $ the Einstein equations $\left(
\ref{Einstein_rearranged_1}\right)  $ become%
\begin{equation}
K^{\mu}R_{\mu\nu}=\tfrac{1}{3}\nabla_{\rho}\left(  2Q_{IJ}\lambda^{I}F_{\nu
}^{J\text{ }\rho}+Q^{IJ}\Lambda_{I\sigma}G_{J\nu}^{\text{ \ \ }\rho\sigma
}\right)  +\tfrac{1}{6}Q^{IJ}H_{I}^{\left(  2\right)  \rho\sigma}G_{J\nu
\rho\sigma}. \label{Ricci_Killing}%
\end{equation}

From this follows the Komar mass integral
\cite{Gibbons:1993xt,Sabra:1997yd,Myers:1986un,Peet:2000hn,Gibbons:2013tqa}
over the spatial hypersurface, $\Sigma_{4}$, including the boundary terms over
$X^{3}$\footnote{The here used convention $\operatorname{div}X=-\delta
X\Rightarrow\delta dZ_{I}=-\hat{\nabla}^{2}Z_{I}$, where $\delta$ is the to
$d$ adjoint exterior derivative, means for here: $\nabla^{2}K_{\mu}=R_{\mu\nu
}K^{\nu}$, so the opposite sign as in \cite{Gibbons:2013tqa}.}:%
\begin{equation}
M=-\tfrac{1}{16\pi G_{5}}\left[  \int_{\Sigma_{4}}H_{I}^{\left(  2\right)
}\wedge F^{I}-\int_{X^{3}}\left(  2\lambda^{I}G_{I}-\Lambda_{I}\wedge
F^{I}\right)  \right]  .\label{Komar mass}%
\end{equation}

Since in \cite{Gibbons:2013tqa} the spacetime was assumed to be asymptotic to
$%
%TCIMACRO{\U{211d} }%
%BeginExpansion
\mathbb{R}
%EndExpansion
^{1,4}$, the boundary integral was taken over $X^{3}=S^{3}$. Here the
spacetime will be asymptotic to $%
%TCIMACRO{\U{211d} }%
%BeginExpansion
\mathbb{R}
%EndExpansion
^{1,3}\times S^{1}$ and so we have $X^{3}=S_{\infty}^{2}\times S^{1}$.

\subsection{Spacetime with a \textquotedblleft running bolt\textquotedblright%
\ and conserved charges}

The five-dimensional metric, called the \textquotedblleft running
bolt\textquotedblright\ \cite{Warner:2009} is a time fibration over Euclidian
Schwarzschild:%
\begin{align}
ds_{5}^{2} &  =-Z^{-2}\left(  dt+k\right)  +Zds_{4}^{2}\nonumber\\
&  =-Z^{-2}\left(  dt+k\right)  +Z\left[  \left(  1-\tfrac{2m}{r}\right)
d\tau^{2}+\left(  1-\tfrac{2m}{r}\right)  ^{-1}dr^{2}+r^{2}\left(  d\theta
^{2}+\sin^{2}\theta d\phi^{2}\right)  \right]  .
\end{align}
The Maxwell fields are set up by the \textquotedblleft floating
brane\textquotedblright\ ansatz \cite{FloatingBranes2010},%
\begin{equation}
A^{I}=-\varepsilon Z_{I}^{-1}\left(  dt+k\right)  +B^{\left(  I\right)
},\label{A}%
\end{equation}
where $\varepsilon$ is set by the (anti-)self-duality of the fields. The
magnetic field strengths are%
\begin{equation}
\Theta^{\left(  I\right)  }=dB^{\left(  I\right)  }.\label{Theta}%
\end{equation}
The three forms, $Z_{I}$, $\Theta^{\left(  I\right)  }$ and $k$, are
determined through the equations
\cite{GutowskiReall,BenaWarner2005,GoldsteinKatmadas,Warner:2009}:%
\begin{align}
\Theta^{\left(  I\right)  } &  =\varepsilon\star_{4}\Theta^{\left(  I\right)
},\label{base1}\\
\hat{\nabla}^{2}Z_{I} &  =\tfrac{1}{2}\varepsilon C_{IJK}\star_{4}\left[
\Theta^{\left(  J\right)  }\wedge\Theta^{\left(  K\right)  }\right]
,\label{base2}\\
dk+\varepsilon\star_{4}dk &  =\varepsilon Z_{I}\Theta^{\left(  I\right)
}.\label{base3}%
\end{align}
Note, that $\left(  \ref{base1}\right)  -\left(  \ref{base3}\right)  $ are
purely represented on the base manifold.

Following the choice of solution for the field strength made in
\cite{Warner:2009},%
\begin{equation}
\Theta^{\left(  I\right)  }=q_{I}\left(  \tfrac{1}{r^{2}}d\tau\wedge
dr+\varepsilon d\Omega_{2}\right)  , \label{Field strength}%
\end{equation}
we have also%
\begin{align}
Z_{I}  &  =1-\tfrac{1}{2m}\tfrac{1}{r}C_{IJK}q_{J}q_{K}\label{Z}\\
k  &  =\mu\left(  r\right)  d\tau=\varepsilon\left(  \tfrac{1}{r}-\tfrac
{1}{2m}\right)  \left[  \Sigma_{I=1}^{3}q_{I}-\tfrac{3}{2m}q_{1}q_{2}%
q_{3}\left(  \tfrac{1}{r}+\tfrac{1}{2m}\right)  \right]  d\tau, \label{k}%
\end{align}
where the $q_{I}$ are $M5$-charges associated with the magnetic field strength component.

It is important to note that harmonic terms $\propto d\tau$ have been chosen
such that $k$ vanishes on the bolt, which is essential to remove closed
timelike curves. With this choice, the asymptotic limit of the angular
momentum does not vanish but has a finite value:%
\begin{equation}
\mu\overset{r\rightarrow\infty}{\rightarrow}\gamma=-\tfrac{\varepsilon}%
{2m}\left(  \Sigma_{I=1}^{3}q_{I}-\tfrac{3}{4m^{2}}q_{1}q_{2}q_{3}\right)  .
\label{gamma}%
\end{equation}
It is this finite limit which led to the name \textquotedblleft running
bolt\textquotedblright.

Transforming $\left(  \ref{gamma}\right)  $ leads to a formula for the
magnetic charges:%

\begin{equation}
\Sigma_{I=1}^{3}q_{I}=-2\varepsilon m\gamma+\tfrac{3}{4m^{2}}q_{1}q_{2}q_{3}.
\label{M5-charge}%
\end{equation}

Equations $\left(  \ref{dG}\right)  $, $\left(  \ref{G_5D}\right)  $ and
$\left(  \ref{Metric_kinetic terms}\right)  $, lead to the conserved
charge-densities:%
\begin{equation}
0=d\left(  2Q_{IJ}\star_{5}dA^{J}-\tfrac{1}{2}C_{IJK}A^{J}\wedge
dA^{K}\right)  .
\end{equation}
Using $\left(  \ref{A}\right)  $, $\left(  \ref{Theta}\right)  $ and $\left(
\ref{Field strength}\right)  -\left(  \ref{k}\right)  $, one can the define
and compute the conserved electric $M2$-charges \cite{Warner:2009},%
\begin{align}
Q^{I}  &  =\int_{S^{1}\times S_{\infty}^{2}}\left(  2G_{I}-\tfrac{1}{2}%
C_{IJK}A^{J}\wedge dA^{K}\right) \label{M2-charge_integral}\\
&  =-\left(  8\pi m\right)  \left(  4\pi\right)  \tfrac{1}{2}C_{IJK}\left[
\tfrac{\varepsilon}{m}q_{J}q_{K}+\tfrac{\gamma}{2}\left(  q_{J}+q_{K}\right)
\right]  . \label{M2-charge}%
\end{align}

\section{Komar mass in five-dimensional almost-BPS supergravity}

\subsection{Setting up the mass formula}

The generic timelike Killing vector may be equipped with an extra $\tau
$-component associated with the angular momentum, so near infinity we can
write:%
\begin{equation}
K=\tfrac{\partial}{\partial t}+\alpha\tfrac{\partial}{\partial\tau},
\label{Killing_general}%
\end{equation}
where $\alpha$ is a constant.

In the following we derive expressions for the fields and fluxes from the RHS
of $\left(  \ref{Komar mass}\right)  $ to understand their contributations to
the mass formula.

From $\left(  \ref{A}\right)  ,\left(  \ref{Theta}\right)  $ and $\left(
\ref{Field strength}\right)  $ follows the Maxwell-field strength,
$F^{I}=dA^{I}$, which decomposes into an exact and a harmonic part,%
\begin{equation}
F^{I}=dA^{I}+\varepsilon q_{I}d\Omega_{2},\label{F_decomposed}%
\end{equation}
where%
\begin{equation}
A^{I}=-\varepsilon Z_{I}^{-1}\left(  dt+\mu d\tau\right)  +q_{I}\left(
\tfrac{1}{r}-\tfrac{1}{2m}\right)  d\tau.\label{A2}%
\end{equation}
Note that we chose a gauge such that the $A^{\left(  I\right)  }$ vanish at
the bolt and are thus globally smooth.

Now we get from $\left(  \ref{F_G_Killing}a\right)  $:%
\begin{equation}
\lambda^{I}=\left(  1+\alpha\mu\right)  \varepsilon Z_{I}^{-1}-\alpha
\tfrac{q_{I}}{r}-\beta^{I}, \label{lambda_small_general}%
\end{equation}
where $\beta^{I}$ are constants.

With $\left(  \ref{A}\right)  $ and $\left(  \ref{G_5D}\right)  $ we have%
\begin{align}
G_{I}  &  =\tfrac{1}{2}\left[  -\varepsilon r^{2}\left(  1-\tfrac{2m}%
{r}\right)  Z_{I}^{\prime}+\varepsilon r^{2}Z_{I}Z^{-3}\mu\mu^{\prime}%
+q_{I}Z_{I}^{2}Z^{-3}\mu\right]  d\tau\wedge d\Omega_{2}\nonumber\\
&  +\tfrac{1}{2}Z_{I}Z^{-3}\left(  \varepsilon r^{2}\mu^{\prime}+q_{I}%
Z_{I}\right)  dt\wedge d\Omega_{2}+\tfrac{\varepsilon}{2r^{2}}q_{I}Z_{I}%
^{2}Z^{-3}dt\wedge d\tau\wedge dr, \label{G}%
\end{align}
and find from this together with $\left(  \ref{lambda_small_general}\right)
$,%
\begin{align}
i_{K}G_{I}+\tfrac{1}{2}C_{IJK}\lambda^{J}F^{K}  &  =-\tfrac{\alpha\varepsilon
}{4m}C_{IJK}q_{J}q_{K}d\Omega_{2}-\tfrac{1}{2}C_{IJK}\beta^{J}F^{K}\nonumber\\
&  -\tfrac{1}{2}d\left[  Z_{I}Z^{-3}\left(  1+\alpha\mu+\alpha r\mu^{\prime
}+\tfrac{\varepsilon\alpha}{r}q_{I}Z_{I}\right)  dt\right]
\label{Fluxes_combined}\\
&  -\tfrac{1}{2}d\left\{  \left[  \mu Z_{I}Z^{-3}\left(  1+\alpha\mu+\alpha
r\mu^{\prime}+\tfrac{\varepsilon\alpha}{r}q_{I}Z_{I}\right)  +\tfrac{\alpha
C_{IJK}q_{J}q_{K}}{2r^{2}}\right]  d\tau\right\}  ,\nonumber
\end{align}
which is a manifestly closed expression.

Using $\left(  \ref{F_G_Killing}b\right)  $ we can directly read that the
total derivative terms of $\left(  \ref{Fluxes_combined}\right)  $ flow into
$d\Lambda_{I}$, together with all exact pieces of $\left(  \ref{F_decomposed}%
\right)  $, such that we have:%
\begin{align}
\Lambda_{I}  &  =-\tfrac{1}{2}\left\{  Z_{I}Z^{-3}\left(  1+\alpha\mu+\alpha
r\mu^{\prime}+\tfrac{\varepsilon\alpha}{r}q_{I}Z_{I}\right)  \left(  dt+\mu
d\tau\right)  +\tfrac{\alpha C_{IJK}q_{J}q_{K}}{2r^{2}}d\tau\right.
\label{lambda_big_general}\\
&  \left.  +C_{IJK}\beta^{J}\left[  -\varepsilon Z_{K}^{-1}\left(  dt+\mu
d\tau\right)  +\tfrac{1}{r}q_{K}d\tau\right]  \right\}  +\tilde{\Lambda}%
_{I},\nonumber
\end{align}
where $\tilde{\Lambda}_{I}$ is a constant closed 1-form which must be\ fixed
such that $\Lambda_{I}$ is smooth at the bolt:%
\begin{equation}
\tilde{\Lambda}_{I}=\tfrac{1}{4m}C_{IJK}q_{J}\left(  \tfrac{\alpha}{4m}%
q_{K}+\beta^{K}\right)  d\tau+const.\cdot dt. \label{lambda_big_constant}%
\end{equation}
The remaining harmonic term in $\left(  \ref{Fluxes_combined}\right)  $, along
with the nontrivial piece in $\left(  \ref{F_decomposed}\right)  $, sum up to
the 2-form harmonic:%
\begin{equation}
H_{I}^{\left(  2\right)  }=-\tfrac{\varepsilon}{2}C_{IJK}q_{J}\left(
\tfrac{\alpha}{2m}q_{K}+\beta^{K}\right)  d\Omega_{2}. \label{H2_general}%
\end{equation}

The harmonic term is (up to a coefficient) completely given by the bolt's
volume, like the harmonic part of the field strengths. Hence, with $\left(
\ref{F_decomposed}\right)  $ and $\left(  \ref{H2_general}\right)  $\ we see
directly that the bulk integral in eq. $\left(  \ref{Komar mass}\right)  $
becomes%
\begin{align}
\int_{\Sigma_{4}}H_{I}^{\left(  2\right)  }\wedge F^{I}  &  =\int_{\Sigma_{4}%
}H_{I}^{\left(  2\right)  }\wedge dA^{I}=\int_{\Sigma_{4}}d\left(
H_{I}^{\left(  2\right)  }\wedge A^{I}\right)  =\int_{S^{1}\times S_{\infty
}^{2}}H_{I}^{\left(  2\right)  }\wedge A^{I}\nonumber\\
&  =\tfrac{\varepsilon}{2}C_{IJK}q_{J}\left(  \tfrac{\alpha}{2m}q_{K}%
+\beta^{K}\right)  \int_{S^{1}\times S_{\infty}^{2}}\left(  \varepsilon
Z_{I}^{-1}\gamma+\tfrac{1}{2m}q_{I}\right)  d\tau\wedge d\Omega_{2}.
\label{mass_bulk}%
\end{align}
Put in other words, the homology of our base space has no self-intersection,
as opposed to the Gibbons-Hawking base which is elaborated in detail in
\cite{Gibbons:2013tqa}. Because the only non-trivial topology lies in the
volume of the $S^{2}$ and so has canceled out in the bulk term, the mass in
the present case is a pure boundary integral.

\subsection{The asymptotic mass formula}

In this section we will evaluate $\left(  \ref{Komar mass}\right)  $. As noted
above, it is a pure boundary integral, so it is convenient to first derive the
asymptotic expressions for the potentials and fields to see what is left of
each term at infinity.

From $\left(  \ref{F_decomposed}\right)  -\left(  \ref{G}\right)  $ and
$\left(  \ref{lambda_big_general}\right)  -\left(  \ref{lambda_big_constant}%
\right)  $\ follows:%
\begin{align}
\lambda^{I} &  \rightarrow\left(  1+\alpha\gamma\right)  \varepsilon-\beta
^{I},\label{lambda_small}\\
\Lambda_{I} &  \rightarrow-\tfrac{1}{2}\left[  \left(  1+\alpha\gamma\right)
\gamma-\tfrac{\alpha}{8m^{2}}C_{IJK}q_{J}q_{K}-\Sigma_{J=1}^{3}C_{IJK}\left(
\varepsilon\gamma+\tfrac{1}{2m}q_{J}\right)  \beta^{K}\right]  d\tau
+...,\label{lambda_big}\\
F^{I} &  \rightarrow\varepsilon q_{I}d\Omega_{2}+...,\label{F_asymptotic}\\
G_{I} &  \rightarrow\tfrac{1}{2}\left(  2\varepsilon m\gamma^{2}-\tfrac
{3}{4m^{2}}\gamma q_{1}q_{2}q_{3}-\tfrac{\varepsilon}{2m}C_{IJK}q_{J}%
q_{K}+q_{I}\gamma\right)  d\tau\wedge d\Omega_{2}+...,\label{G_asymptotic}%
\end{align}
where the remaining terms don't contribute to the boundary's volume form,
$dvol\left(  S^{1}\times S^{2}\right)  \propto d\tau\wedge d\Omega_{2}$.

Plugging $\left(  \ref{mass_bulk}\right)  $ into $\left(  \ref{Komar mass}%
\right)  $ we find with $\left(  \ref{lambda_small}\right)  -\left(
\ref{G_asymptotic}\right)  $ the purely boundary induced mass:%
\begin{align}
M  &  =\tfrac{\pi}{2G_{5}}\left[  12m^{2}\gamma^{2}\left(  1+\alpha
\gamma\right)  -\left(  3\alpha\gamma+2\right)  \Sigma_{I=1}^{3}C_{IJK}%
q_{J}q_{K}-\tfrac{9\varepsilon}{2m}\left(  \alpha\gamma^{2}+\gamma
+\alpha\right)  q_{1}q_{2}q_{3}\right] \nonumber\\
&  =\tfrac{2\pi}{G_{5}}\left(  \tfrac{2+3\alpha\gamma}{64\pi^{2}}%
\varepsilon\Sigma_{I=1}^{3}Q^{I}-\tfrac{3\alpha+\gamma}{2}\varepsilon
m\Sigma_{I=1}^{3}q_{I}-3\alpha\gamma m^{2}\right)  . \label{Komar mass_2}%
\end{align}
Note that the complete $\beta^{I}$-dependence has canceled out, as it must for
the mass to be gauge-invariant.

The above result could have also been achieved by directly calculating%
\begin{equation}
M=\tfrac{3}{32\pi G_{5}}\int_{S^{1}\times S_{\infty}^{2}}\star_{5}dK=\tfrac
{3}{32\pi G_{5}}\int_{S^{1}\times S_{\infty}^{2}}\star_{5}\left[  \partial
_{r}\left(  g_{00}+\alpha g_{01}\right)  dr\wedge dt+\partial_{r}\left(
g_{01}+\alpha g_{11}\right)  dr\wedge d\tau\right]  .
\end{equation}
It is very helpful to use the frames of the five-dimensional metric,%
\begin{equation}%
\begin{array}
[c]{ll}%
e^{0}=Z^{-1}\left(  dt+\mu d\tau\right)  & e^{1}=Z^{\frac{1}{2}}\left(
1-\frac{2m}{r}\right)  ^{\frac{1}{2}}d\tau\\
e^{2}=Z^{\frac{1}{2}}\left(  1-\frac{2m}{r}\right)  ^{-\frac{1}{2}}dr &
e^{3}\wedge e^{4}=Zr^{2}d\Omega_{2}%
\end{array}
,
\end{equation}
to easily compute the duals in the above integral.

Now we consider two natural special cases for the Killing vector -- the rest
frame, $K=\frac{\partial}{\partial t}$, and the asymptotically static frame,
$K=\frac{\partial}{\partial\hat{t}}$, where the latter is based upon the at
infinity co-rotating coordinate system with time-coordinate, $\hat{t}$.

For that purpose, it is convenient to first write a more general form of the
Killing vector:%
\begin{equation}
K=\alpha_{0}\tfrac{\partial}{\partial t}+\alpha_{1}\tfrac{\partial}%
{\partial\tau}. \label{K}%
\end{equation}
In this form eq. $\left(  \ref{Komar mass_2}\right)  $ becomes with $\left(
\ref{M5-charge}\right)  $ and $\left(  \ref{M2-charge}\right)  $:%
\begin{equation}
M=\tfrac{2\pi}{G_{5}}\left(  \tfrac{2\alpha_{0}+3\alpha_{1}\gamma}{64\pi^{2}%
}\varepsilon\Sigma_{I=1}^{3}Q^{I}-\tfrac{3\alpha_{1}+\gamma\alpha_{0}}%
{2}\varepsilon m\Sigma_{I=1}^{3}q_{I}-3\alpha_{1}\gamma m^{2}\right)  .
\label{Komar mass_more general}%
\end{equation}

In order to get the mass in the asymptotically static frame one has to
consider the coordinates given in \cite{Warner:2009}, eqs. (3.20), and choose
the Killing vector accordingly, $K=\tfrac{\partial}{\partial\hat{t}}%
=\sqrt{1-\gamma^{2}}\tfrac{\partial}{\partial t}+\tfrac{\gamma}{\sqrt
{1-\gamma^{2}}}\tfrac{\partial}{\partial\tau}$:%
\begin{equation}
M_{a.s.}=\tfrac{2\pi}{G_{5}\sqrt{1-\gamma^{2}}}\left[  \tfrac{\gamma^{2}%
+2}{64\pi^{2}}\varepsilon\Sigma_{I=1}^{3}Q^{I}+\tfrac{\gamma\left(  \gamma
^{2}-4\right)  }{2}\varepsilon m\Sigma_{I=1}^{3}q_{I}-3m^{2}\gamma^{2}\right]
. \label{Mass_asymptotically static}%
\end{equation}
This one will become handy for the later discussion of the mass terms.

To go to the rest frame, one has to take $K=\tfrac{\partial}{\partial t}$,
that is, choose $\alpha_{0}=1$ and $\alpha_{1}=0$:%
\begin{equation}
M_{0}=\tfrac{\pi}{4G_{5}}\varepsilon\left(  \tfrac{1}{4\pi^{2}}\Sigma
_{I=1}^{3}Q^{I}-4m\gamma\Sigma_{I=1}^{3}q_{I}\right)  . \label{Mass_rest}%
\end{equation}
We see already here that the BPS-bound break is due to the magnetic
$M5$-charges which are driving the bolt; a closer analysis follows in the next subsection.

\subsection{Mass term analysis}

In the following we want to illuminate the explicit origin of the extra mass
term which violates the BPS-bound. For this purpose, we will substitute the
total Maxwell-charge into the mass formula and consider the remaining terms in detail.

We work in the rest frame and set $\alpha=0$.

With the identity,%
\begin{equation}
F\wedge F=\left(  dA\right)  ^{\wedge2}+dA\wedge F_{\text{harmonic}%
}+F_{\text{harmonic}}\wedge dA+F_{\text{harmonic}}^{\wedge2}%
=2F_{\text{harmonic}}\wedge dA,
\end{equation}
which follows from the fact that $\left(  dA\right)  ^{\wedge2}\propto\left(
dr\wedge d\tau\right)  ^{\wedge2}=0$ and $F_{\text{harmonic}}^{\wedge2}%
\propto\left(  d\Omega_{2}\right)  ^{\wedge2}=0$ as a direct consequence of
the non-intersecting homology, and $\left(  \ref{F_asymptotic}\right)  $ and
$\left(  \ref{dG}\right)  $ one can rewrite $\left(  \ref{mass_bulk}\right)  $
like%
\begin{align}
-\int_{\Sigma_{4}}H_{I}^{\left(  2\right)  }\wedge F^{I}  &  =\tfrac
{\varepsilon}{2}C_{IJK}\beta^{I}q_{J}\int_{\Sigma_{4}}d\Omega_{2}\wedge
dA^{K}=\tfrac{1}{2}C_{IJK}\beta^{I}\int_{\Sigma_{4}}F_{\text{harmonic}}%
^{J}\wedge dA^{K}\nonumber\\
&  =\tfrac{1}{4}C_{IJK}\beta^{I}\int_{\Sigma_{4}}F^{J}\wedge F^{K}=\beta
^{I}\int_{\Sigma_{4}}dG_{I}=\beta^{I}\int_{S^{1}\times S_{\infty}^{2}}G_{I}.
\end{align}
Now, with this, $\left(  \ref{lambda_small}\right)  $ and $\left(
\ref{M2-charge_integral}\right)  $ eq. $\left(  \ref{Komar mass}\right)  $
becomes%
\begin{align}
M_{0}  &  =\tfrac{1}{16\pi G_{5}}\left\{  \beta^{I}\int_{S^{1}\times
S_{\infty}^{2}}G_{I}+\int_{S^{1}\times S_{\infty}^{2}}\left[  2\Sigma
_{I=1}^{3}\left(  \varepsilon-\beta^{I}\right)  G_{I}-\Lambda_{I}\wedge
F^{I}\right]  \right\} \nonumber\\
&  =\tfrac{1}{16\pi G_{5}}\int_{S^{1}\times S_{\infty}^{2}}\left[
\Sigma_{I=1}^{3}\left(  2\varepsilon-\beta^{I}\right)  G_{I}-\Lambda_{I}\wedge
F^{I}\right] \label{Mass_rest_rewritten}\\
&  =\tfrac{1}{16\pi G_{5}}\left\{  \varepsilon\Sigma_{I=1}^{3}Q^{I}%
+\int_{S^{1}\times S_{\infty}^{2}}\left[  \tfrac{\varepsilon}{2}\Sigma
_{I=1}^{3}C_{IJK}A^{J}\wedge F^{K}-\beta^{I}G_{I}-\Lambda_{I}\wedge
F^{I}\right]  \right\}  ,\nonumber
\end{align}
where the $\beta$-term cancels with the $\beta$-term inherent to $\Lambda
_{I}\wedge F^{I}$ (see $\left(  \ref{lambda_big}\right)  $).

There will be no contribution by $G_{I}$ left, if we choose $\beta^{I}=0$; if,
on the other hand, we say $\beta^{I}=\beta=\tfrac{2m\gamma\Sigma_{I}q_{I}%
}{4m\varepsilon\gamma\Sigma_{I}q_{I}+\Sigma_{I}C_{IJK}q_{J}q_{K}}$, then
$\Lambda_{I}\wedge F^{I}$ vanishes from the integral. Hence, the degrees of
freedom, $\beta^{I}$, allow one to move the contributions of these two terms
around, which sum up to%
\begin{equation}
-\tfrac{1}{16\pi G_{5}}\int_{S^{1}\times S^{2}}\left(  \beta^{I}G_{I}%
+\Lambda_{I}\wedge F^{I}\right)  =\tfrac{\pi m}{G_{5}}\varepsilon\gamma
\Sigma_{I=1}^{3}q_{I}. \label{BPS_break_lambda_F}%
\end{equation}
Furthermore, the substitution of the total Maxwell-charge into the mass
formula brought up another boundary term:%
\begin{equation}
\tfrac{1}{16\pi G_{5}}\tfrac{\varepsilon}{2}\Sigma_{I=1}^{3}C_{IJK}\int
_{S^{1}\times S^{2}}A^{J}\wedge F^{K}=-\tfrac{2\pi m}{G_{5}}\varepsilon
\gamma\Sigma_{I=1}^{3}q_{I}. \label{BPS_break_M2_part}%
\end{equation}

The contributions to the BPS-bound violation by $\left(
\ref{BPS_break_lambda_F}\right)  $ and $\left(  \ref{BPS_break_M2_part}%
\right)  $ are obviously opposed to each other.

Note: One could also choose $\beta^{I}$ to make both the $A\wedge F$ term and
the $\Lambda\wedge F$ term vanish and so to turn the extra-mass into a pure
integral over $G_{I}$,%
\begin{equation}
\beta^{I}=\beta=\varepsilon\left(  \tfrac{2m\gamma\Sigma_{I}^{3}q_{I}%
}{4m\gamma\Sigma_{I}q_{I}+\varepsilon\Sigma_{I=1}^{3}C_{IJK}q_{J}q_{K}%
}-1\right)  ,
\end{equation}
in which case%
\begin{equation}
M_{0}=\tfrac{1}{16\pi G_{5}}\left(  \varepsilon\Sigma_{I=1}^{3}Q^{I}%
-\int_{S^{1}\times S_{\infty}^{2}}\beta^{I}G_{I}\right)  .
\end{equation}

In any case, one can see that the main agents are the field strengths, $F^{I}$
and $G_{I}$: They do not fall off towards infinity but have according to
$\left(  \ref{F_asymptotic}\right)  $ and $\left(  \ref{G_asymptotic}\right)
$\ \textquotedblleft surviving\textquotedblright\ legs in the bolt's volume,
$F^{I}\rightarrow\varepsilon q_{I}d\Omega_{2}$, and the whole boundary's
volume, $G_{I}\rightarrow\left[  ...\right]  d\tau\wedge d\Omega_{2}$, respectively.

\section{Relation between the five- and the four-dimensional mass}

The difference between the above derived masses is:%
\begin{equation}
M_{0}=\tfrac{1}{\sqrt{1-\gamma^{2}}}\left(  M_{a.s.}-\tfrac{3}{2}\gamma
Q_{e}\right)  , \label{mass_difference}%
\end{equation}
where $Q_{e}$ is the Kaluza-Klein charge,%
\begin{equation}
Q_{e}=-\tfrac{\pi}{G_{5}\sqrt{1-\gamma^{2}}}\left[  \tfrac{3\varepsilon}%
{2m}\left(  1+\gamma^{2}\right)  q_{1}q_{2}q_{3}+\gamma\Sigma_{I=1}^{3}%
C_{IJK}q_{J}q_{K}-4m^{2}\gamma^{3}\right]  .
\end{equation}

So, the asymptotically static mass, $M_{a.s.}$, is related to the in the rest
frame moving mass, $M_{0}$, by the usual relativistic factor, $\sqrt
{1-\gamma^{2}}$, and an additional shift by the Kaluza-Klein charge induced by
the bolt\footnote{See eqs. (3.26)-(3.27) of \cite{Warner:2009}, where the
KK-gauge field is read off from the metric in the asymptotically static frame
as the $\hat{t}$-shift of the $\hat{\tau}$-circle.}.

\subsection{Relating the masses}

The mass of the solution found in \cite{Warner:2009} was optained after
dimensional reduction to four dimensions, whereas here we have considered the
intrinsic five-dimensional Komar mass. In the following we examine the
relation between these in more detail.

The dimensional reduction in \cite{Warner:2009} happened in the asymptotically
static frame along the $\hat{\tau}$-circle. The asymptotic mass was then read
off from the $\left(  0,0\right)  $-coefficient of the emergent
four-dimensional Einstein-metric. We denote this mass by $M_{a.s.}^{\left(
4\right)  }$ to distinguish it from the asymptotically static mass, $M_{a.s.}%
$, derived earlier.

This mass can also be optained from a four-dimensional Komar integral:%
\begin{equation}
M_{a.s.}^{\left(  4\right)  }=-\tfrac{1}{8\pi G_{4}}\int_{S^{2}}\star
_{4}dK_{E}, \label{Mass_4D}%
\end{equation}
where $E$ denotes the connection to the four-dimensional Einstein-metric,
$K_{E}=g_{00}^{E}d\hat{t}$, and the normalization is received in the same
manner as for $M_{a.s.}$ with the different assumption of the mass density
referring to a three-dimensional volume, $M_{a.s.}^{\left(  4\right)  }%
=\int_{\Sigma_{3}}T_{00}^{E}d\Sigma_{3}$.

It is useful to show the explicit relation between the five-dimensional metric
and the four-dimensional Einstein-metric in order to deduct the relation
between the masses. This can be optained from eq. $\left(  3.21\right)  $ of
\cite{Warner:2009},%
\begin{align}
ds_{5}^{2}  &  =w^{-2}\left(  d\hat{\tau}+\omega d\hat{t}\right)  ^{2}%
+wds_{E}^{2},\label{Metric_relation}\\
ds_{E}^{2}  &  =-\hat{I}_{4}^{-\frac{1}{2}}d\hat{t}^{2}+\hat{I}_{4}^{\frac
{1}{2}}\left[  dr^{2}+\left(  1-\tfrac{2m}{r}\right)  r^{2}\left(  d\theta
^{2}+\sin^{2}\theta d\phi^{2}\right)  \right] \nonumber
\end{align}
with the warp factors,%
\begin{equation}
w=Z\left[  \tfrac{1-\gamma^{2}}{Z^{3}\left(  1-\frac{2m}{r}\right)  -\mu^{2}%
}\right]  ^{\frac{1}{2}}\text{ and }\hat{I}_{4}^{\frac{1}{2}}=-\left(
g_{00}^{E}\right)  ^{-1}=\tfrac{1}{1-\frac{2m}{r}}\sqrt{\tfrac{Z^{3}\left(
1-\frac{2m}{r}\right)  -\mu^{2}}{1-\gamma^{2}}}, \label{Metric_components}%
\end{equation}
and%
\begin{equation}
\omega=w^{2}g_{01}=\gamma-\left(  1-\tfrac{2m}{r}\right)  ^{-2}\mu\hat{I}%
_{4}^{-1}.
\end{equation}
Furthermore, the two sets of frames are connected like%
\begin{equation}%
\begin{array}
[c]{lllll}%
\text{five-dimensional:} & e^{0}=\sqrt{w}e_{E}^{0} & e^{2}=\sqrt{w}e_{E}^{2} &
e^{3}\wedge e^{4}=we_{E}^{3}\wedge e_{E}^{4} & e^{1}=w^{-1}\left(  d\hat{\tau
}+\omega d\hat{t}\right) \\
\text{four-dimensional:} & e_{E}^{0}=\hat{I}_{4}^{-\frac{1}{4}}d\hat{t} &
e_{E}^{2}=\hat{I}_{4}^{\frac{1}{4}}dr & e_{E}^{3}\wedge e_{E}^{4}=\hat{I}%
_{4}^{\frac{1}{2}}\left(  1-\frac{2m}{r}\right)  r^{2}d\Omega_{2} &
\end{array}
\end{equation}

In order to derive the relation between the masses, we take a closer look at
the Killing-vectors, $K$ and $K_{E}$; in the asymptotically static frame they
are both equal to $\frac{\partial}{\partial\hat{t}}$, but the associated
one-forms result from lowering with the different metrics,%
\begin{equation}
K=g_{0\mu}d\hat{x}^{\mu}=g_{00}d\hat{t}+g_{01}d\hat{\tau}\text{ and }%
K_{E}=g_{0\mu}^{E}d\hat{x}^{\mu}=g_{00}^{E}d\hat{t}.
\end{equation}
Working with frames, one can easily show that%
\begin{align}
\star_{5}dK  &  =\star_{5}\left(  \partial_{r}g_{00}dr\wedge d\hat{t}%
+\partial_{r}g_{01}dr\wedge d\hat{\tau}\right) \nonumber\\
&  =w^{-1}\left(  \partial_{r}g_{00}-\omega\partial_{r}g_{01}\right)
\star_{5}\left(  e^{2}\wedge e^{0}\right)  +\hat{I}_{4}^{-\frac{1}{4}}%
w^{\frac{1}{2}}\partial_{r}g_{01}\star_{5}\left(  e^{2}\wedge e^{1}\right)
\nonumber\\
&  =-w^{-1}\left(  \partial_{r}g_{00}-\omega\partial_{r}g_{01}\right)
e^{1}\wedge e^{3}\wedge e^{4}-\hat{I}_{4}^{-\frac{1}{4}}w^{\frac{1}{2}%
}\partial_{r}g_{01}e^{0}\wedge e^{3}\wedge e^{4}\\
&  =-2\hat{I}_{4}^{\frac{3}{2}}\left(  w\partial_{r}\hat{I}_{4}\right)
^{-1}\left(  \partial_{r}g_{00}-\omega\partial_{r}g_{01}\right)  d\hat{\tau
}\wedge\star_{4}dK_{E}+\text{\textquotedblleft terms with }d\hat
{t}\text{\textquotedblright.}\nonumber
\end{align}
Integrating the last equation as in $\left(  \ref{Komar integral}\right)  $
yields%
\begin{align}
M_{a.s.}  &  =\tfrac{3}{32\pi G_{5}}\int_{S^{1}\times S_{\infty}^{2}}\star
_{5}dK=-\tfrac{3vol\left(  S^{1}\right)  }{16\pi G_{5}}\tfrac{\partial
_{r}g_{00}}{\partial_{r}\hat{I}_{4}}|_{r\rightarrow\infty}\int_{S_{\infty}%
^{2}}\star_{4}dK_{E}\nonumber\\
&  =\tfrac{3}{4}\tfrac{\partial_{r}g_{00}}{\partial_{r}g_{00}^{E}%
}|_{r\rightarrow\infty}M_{a.s.}^{\left(  4\right)  }=\tfrac{3}{4}\left(
1-\tfrac{2\partial_{r}w}{\partial_{r}\hat{I}_{4}}|_{r\rightarrow\infty
}\right)  M_{a.s.}^{\left(  4\right)  }, \label{M5_M4}%
\end{align}
where we have used that the gravitational constants are linked like%
\begin{equation}
G_{4}=\tfrac{1}{vol\left(  S^{1}\right)  }G_{5}=\tfrac{1}{8\pi m\sqrt
{1-\gamma^{2}}}G_{5}. \label{Gravititational constants}%
\end{equation}
In any case, we see that the masses are proportional.

One can show that for a special choice of $\alpha_{0}$ and $\alpha_{1}$ in
$\left(  \ref{K}\right)  $ the masses match, but it involves unintuitive
ratios of the derivatives of the metrics and will not be further elaborated at
this point.

\subsection{Smarr formula for a four-dimensional mass through
\textquotedblleft dimensional extension\textquotedblright}

If, on the other hand, one has with a four-dimensional spacetime and would
like to express the asymptotic mass by means of an integral over topology in
the sense of a Smarr formula, then one would face the problem of
singularities. However, one could circumvent that issue by \textquotedblleft
dimensional extension\textquotedblright, that is, assuming a curled up
extra-dimension -- like the $S^{1}$ -- and use the inverse form of $\left(
\ref{M5_M4}\right)  $ to write a bit more generally:%
\begin{equation}
M_{a.s}^{\left(  4\right)  }=\zeta_{E}\int_{S_{\infty}^{2}}\star_{4}%
dK_{E}=\tfrac{\zeta_{E}}{vol\left(  S^{1}\right)  }\int_{S^{1}\times
S_{\infty}^{2}}d\hat{\tau}\wedge\star_{4}dK_{E}=-\left(  1-\tfrac{\partial
_{r}w}{\partial_{r}g_{00}^{E}}|_{r\rightarrow\infty}\right)  ^{-1}\tfrac
{\zeta_{E}}{vol\left(  S^{1}\right)  }\int_{S^{1}\times S_{\infty}^{2}}%
\star_{5}dK, \label{M4_M5}%
\end{equation}
where $\zeta_{E}$ is the four-dimensional normalization and $vol\left(
S^{1}\right)  $ the extra-dimension's volume in the asymptotically static
frame. (In the situation above they are $\zeta_{E}=-\tfrac{1}{8\pi G_{4}}$ and
$vol\left(  S^{1}\right)  =8\pi m\sqrt{1-\gamma^{2}}$). The warp-factor, $w$,
results from embedding the four-dimensional into the five-dimensional metric,
as in $\left(  \ref{Metric_relation}\right)  $.

Now, with the five-dimensional Komar integral in $\left(  \ref{M4_M5}\right)
$ one can work in the same manner as done in subsection 3.1 to set up a
five-dimensional Smarr formula for the four-dimensional asymptotic mass.

\section{Conclusion}

We have derived the Komar mass for an almost-BPS solution of supergravity in a
five-dimensional stationary spacetime where we gave space a \textquotedblleft
bolt\textquotedblright\ at the center and made it asymptotically $S^{1}\times%
%TCIMACRO{\U{211d} }%
%BeginExpansion
\mathbb{R}
%EndExpansion
^{3}$.

The very goal was to determine explicitely how each mass component follows
from topology and especially which field and flux terms account particularly
for the extra-mass causing the violation of the BPS-bound.

At first, the whole mass formula turned out to be a pure boundary integral.
This is due to the fact that the only harmonic form in the present spacetime
is the volume form of the bolt, $d\Omega_{2}$, which squares to zero, and so
the wedge-product term in the bulk integral leaves only exact pieces. In other
words, the topology of the base space does not inhabit any self-intersecting
homology as opposed to the Gibbons-Hawking base.

It could be shown that, how much each field and flux term contributes to the
extra-mass, is variable by a degree of freedom; one could, for example, gauge
such that the extra-mass becomes a boundary integral solely over the dual
field strengths, $G_{I}$. In general, the other terms involve the field
strengths, $F^{I}$, whose common harmonic part in the bolt's volume is finite
at infinity and so gives nonvanishing terms in the boundary integral.

Furthermore, the deviation between the masses obtained here and in
\cite{Warner:2009} could be explained by the fact that different-dimensional
spacetimes underlie them. The mass computed in this work ensued from the Komar
integral in the present five dimensions, but the one from \cite{Warner:2009}
is footed on the four-dimensional Einstein-metric of the along the $S^{1}$
reduced spacetime. This specific nature of the latter mass was explicitely
pointed out by writing it in terms of a four-dimensional Komar-integral. Based
on this, a formula was set up to relate these masses.

Also, this formula was shown to be invertible to a formalism allowing the mass
of a four-dimensional spacetime to be written in terms of a five-dimensional
Smarr formula and hence an integral over topology without
singularities.\bigskip\bigskip

\leftline{\bf Acknowledgements}First I would like to express my gratitude to
Nicholas Warner for his great work and support as my doctoral adviser. I am
also thankful for friendly cooperation of the High Energy Group of USC and the
relaxing, work-supportive atmosphere of the University of Southern California.
Last but not least would I like to thank DOE grant DE-SC0011687 by whom this
work was in part supported.

\end{document}